**Proximity-induced giant spin-orbit interaction in epitaxial graphene on topological insulator**

*Kyung-Hwan Jin[1] and Seung-Hoon Jhi[1,2]*

[1]Department of Physics, Pohang University of Science and Technology, Pohang 790-784, Republic of Korea

[2]Division of Advanced Materials Science, Pohang University of Science and Technology, Pohang 790-784, Republic of Korea

Abstract

Heterostructures of Dirac materials such as graphene and topological insulators provide interesting platforms to explore exotic quantum states of electrons in solids. Here we study the electronic structure of graphene-$Sb_2Te_3$ heterostructure using density functional theory and tight-binding methods. We show that the epitaxial graphene on $Sb_2Te_3$ turns into quantum spin-Hall phase due to its proximity to the topological insulating $Sb_2Te_3$. It is found that the epitaxial graphene develops a giant spin-orbit gap of about ~20 meV, which is about three orders of magnitude larger than that of pristine graphene. We discuss the origin of such enhancement of the spin-orbit interaction and possible outcomes of the spin-Hall phase in graphene.

Heterojunctions of materials with different physical properties have served as a basis for finding new physical states and understanding complex phenomena in condensed matter systems. Diffusion of the order parameters by proximity induces a weak order in the non-ordered materials, generating quantum interference effects. For example, suppercurrents and interference effects due to the Josephson tunneling were observed in graphene in contact with superconductors, restoring the weak localization in graphene.[1, 2] Recently topological insulators, which refer to the states of matters with insulating gaps in the bulk and gapless helical states on the surface, have attracted great attention due to their intriguing electronic structures. Dictated by time-reversal symmetry, the helical surface states termed massless Dirac fermions can move without backscattering on the surface of topological insulators. Heterojunctions of materials with different topological orders can thus provide an interesting platform to explore emerging quantum phenomena of Dirac fermions at the interfaces. For example, it was proposed that exotic particles such as the axion, magnetic monopole, and the Majorana fermion can be realized in hybrid structures of topological insulator-superconductor or topological insulator-ferromagnets.[3, 4]

Graphene is a representative Dirac material and has low energy states with pseudo-helicity and linear energy-momentum dispersion originating from the atomic symmetry. It is appropriate to ask what proximity effects can occur in graphene in contact with topological insulators (TIs). Does the strong spin-orbit interaction in TIs affect the electronic structure of graphene? Kane and Mele studied the possibility of spin-Hall phase in graphene by introducing an orbital-symmetry and time-reversal-symmetry preserving term.[5] However, the strength of spin-orbit coupling (SOC) in graphene is extremely small and the spin-Hall phase is expected to occur at very low temperatures of a few Kelvin.[6-9] The intrinsic and Rashba spin-orbit interactions in pristine graphene arise from hybridization between π and σ bands.[6] Enhancing the hybridization, for instance, by adsorbing hydrogen adatoms has been suggested to increase the SOC in graphene [10], or simply adsorbing heavy elements like thallium on graphene was proposed to induce spin-Hall phase in graphene [11]. From other perspectives, direct measurements of transport characteristics of TI surface states, which are crucial for developing the TI devices [12], have been tried after verification of TI surface states by angle-resolved photoemission spectroscopy [13-15]. However, in most TI materials, the Fermi level lies in the conduction bands or valence bands [15, 16] and the bulk conduction dominates over the conduction via the surface states. Ca- or Mn-doping or by intercalation have been tested to align the chemical potential in the middle of TI bulk energy gap [14, 15]. We note that graphene can be an ideal probe to detect TI surface states from the scaling point of view. In this paper, we studied the electronic structure of epitaxial graphene on topological insulating $Sb_2Te_3$ using pseudopotential density functional theory and the tight-binding methods including the spin-orbit interactions. In particular we investigated the proximity effect in the graphene-TI junction and possible spin-Hall phases arising in graphene. By doing so, we also explore graphene-TI hybrid structures as devices to detect the helical surface states.

First-principles calculations based on the density functional theory were carried out using the Vienna ab-initio simulation package.[17] The exchange-correlation interaction of electrons was treated within the generalized gradient approximation (GGA) of Perdew-Burke-Ernzerhof type.[18] Pseudopotentials generated by projector augmented wave method were used for atomic potentials. The SOC was included at the second variational step using the scalar-relativistic eigen-functions as a basis. A cutoff energy of 400 eV was used for the expansion of wave functions and potentials in the plane-wave basis. The $k$-point meshes of 11×11×1 were used for the sampling of the Brillouin zone. For emulating graphene-$Sb_2Te_3$ surface, we used the supercell method by putting a single layer graphene on top of $Sb_2Te_3$ slab and introducing a vacuum layer of 20 Å-thick between the cells to minimize artificial inter-cell interactions. Once full atomic relaxation was done, one additional step of self-consistent calculation was carried out including the SOC until the total energy converges to within $10^{-5}$ eV. Electronic band structures from first-principles calculations were then fit by tight-binding methods including SOC to analyze the origin of energy splitting.

The epitaxial graphene on top of $Sb_2Te_3$ surface was modeled by putting a single layer of graphene on $Sb_2Te_3$ slab of 1 ~ 5 quintuple layers (QLs) with Te atoms at the top [Fig. 1(a)]. We chose the experimental in-plane lattice constant of 4.25 Å for $Sb_2Te_3$ [19] and then adjusted the lattice constant of graphene accordingly. The lattice mismatch by this choice is about ~1% when we used $\sqrt{3}\times\sqrt{3}$ in-plane supercell for graphene. We considered three atomic stacking configurations between graphene and $Sb_2Te_3$ as shown in Fig.1: surface Te atoms at the center of carbon hexagon rings (P1); carbon atoms on top of surface Te atoms (P2); carbon-carbon bridges on top of surface Te atoms (P3). In order to describe the van der Waals-type interaction between graphene and TI surface, we employed a semi-empirical correction by Grimme's method [20] because GGA cannot describe the van der Waals interaction correctly. We found that P1 configuration is the most stable among the three.

Helical surface states of topological insulating phase start to appear over certain thickness of TI slabs. The TI surface states are fully developed in $Sb_2Te_3$ slab of 3QL or thicker, which is common to other topological insulating binary chalcogen compounds such as $Bi_2Se_3$ and $Bi_2Te_3$.[21, 22] Figure 1(b) shows the calculated binding energy curves [$E_b=E_{gra-TI} - (E_g+E_{TI})$; $E_{gra-TI}$, $E_g$, and $E_{TI}$ represent the cohesive energies of graphene-TI (3QL), graphene, and TI (3QL), respectively] with SOC and van der Waals interaction included. The equilibrium binding distance and energy in the P1 configuration are 3.48 Å and about ~41 meV per carbon, respectively. We note that SOC does not affect the binding energy and distance. TI slabs still show a very small but finite energy gap due to the interaction between the surface states at two surfaces of the slabs [21-23]. Figure 1(c) shows our calculated band gaps of $Sb_2Te_3$ as the number of QLs is increased. Without SOC included, the band gap is large for thin slabs due to the quantum confinement effect and then converges to the bulk band gap as the thickness is increased. When SOC is included, the band gap decreases rapidly with increasing slab thickness. Figure 1 (d) shows the electrostatic potential difference ($\Delta V$) in 5QL $Sb_2Te_3$, which represents a change in potential at the surfaces due to the SOC. The electric field by the potential gradient near the surfaces induces the Rashba splitting in graphene.

Now we studied the changes in the graphene electronic structure induced by TI contact. By increasing the slab thickness from 1QL to 4QL, we investigated how emerging TI surface states start to interact with graphene π bands. Our calculated band structures are shown in Fig. 2. A single layer graphene with √3×√3 unit cell should have four-fold degenerate Dirac cones at Γ point due to band folding. On TI substrates (>3QL), we observed a few intriguing features in the graphene Dirac cones; small-gap opening at the Dirac point, splitting in the four-fold degenerate bands particularly in the valence bands, a change in the dispersion of the conduction bands, and the Rashba-type splitting in both the conduction and valence bands. Without SOC, we do not observe such features in the Dirac cone of graphene except the small-gap opening at the Dirac point (see Supplementary Materials Fig. S1). The splitting of the valence bands is of particular interest as its size increases from 25, 41, 47, 52 meV for 1QL, 2QL, 3QL, and 4QL $Sb_2Te_3$, respectively (it is about 53 meV for 5QL). This observation of band splitting along with the Rashba-type splitting indicates that SOC and the inversion symmetry-breaking by TI substrate are playing the major role for the change of the graphene Dirac cones.

In order to understand and resolve the changes in the graphene band structure, we used the tight-binding Hamiltonian of Kane and Mele,[24, 25] that includes both intrinsic and extrinsic SOC terms. The Hamiltonian for 2D honeycomb lattice is given as

$$H = -t\sum_{\langle ij \rangle} c_i^\dagger c_j + \frac{iV_{SO}}{\sqrt{3}} \sum_{\langle\langle ij \rangle\rangle} c_i^\dagger \vec{\sigma} \cdot (\vec{d}_{kj} \times \vec{d}_{ik}) c_j + iV_R \sum_{\langle ij \rangle} c_i^\dagger \hat{e}_z \cdot (\vec{\sigma} \times \vec{d}_{ij}) c_j + iV_{Rh} \sum_{\langle ij \rangle} c_i^\dagger \hat{e}_\rho \cdot (\vec{\sigma} \times \vec{d}_{ij}) c_j$$

The first term is the nearest-neighbor hopping term, and the second term is the intrinsic SOC with a coupling strength $V_{SO}$ with $\vec{\sigma} = (\sigma_x, \sigma_y, \sigma_z)$ being the Pauli matrices, $i$ and $j$ referring next-nearest neighboring sites that have a common nearest neighbor $k$ connected by vectors $\vec{d}_{ik}$ and $\vec{d}_{kj}$. The third and fourth terms are the Rashba SOC due to a electric field normal to the substrate and an in-plane electric field in the substrate, respectively, with $i$ and $j$ referring the nearest neighbors. The parameters for intrinsic spin-orbit, normal and in-plane Rashba spin-orbit interactions are $\lambda_I = 3\sqrt{3}V_{so}/2$, $\Delta_{Rz} = 3V_R/2$, and $\Delta_{Rh} = 3V_{Rh}/2$, respectively. Using this Hamiltonian, we constructed the 12×12 matrix for graphene with a √3×√3 R30° unit cell that has six basis carbon atoms (see Supplementary Materials S.II). The lower panels in Fig. 2 show the band structure from fist-principles calculations fit to the tight binding model. A good agreement of DFT and TB results indicates that the spin-orbit interactions in graphene-TI heterojunction are well represented by the tight-binding model near Γ point. From the fitting to TB model, the band gap opening at the Dirac point is found to originate from a change in the hopping parameter between the next-nearest carbon atoms.

Figure 3 shows the results of intrinsic and extrinsic SOC parameters obtained from fitting first-principles calculations to the TB Hamiltonian as the number of QLs is increased. In all ranges of TI slab thickness, intrinsic SOC strength is much larger than Rashba splitting by both normal and in-plane electric fields. The three parameters ($\lambda_I$, $\Delta_{Rz}$, and $\Delta_{Rh}$) are increased converging to about 20, 8, and 3 meV, respectively, as $Sb_2Te_3$ slab thickness is increased. For other atomic configurations [P2 and P3 in Fig. 1(a)], we found similar results. The SOC strength of about ~20 meV in our calculations is significantly larger than the value of pristine graphene of about 20~50 μeV [6-8] by more than three order of magnitude. This finding of enhanced SOC in graphene by proximity to TI surfaces supports that graphene can work as a probe of the topological surface states by becoming a spin-Hall system.

Along with the spin-orbit gap in the valence bands of about $\sim 2\lambda_I$, we observe some changes in the band dispersion of the conduction bands; the Fermi velocity is decreased to about ≤50% of pristine graphene and cyclotron masses are increased as shown in Fig. 3(b). These changes will affect the electron mobility and the refraction of electron propagation. The spin-helicity of the conduction bands of graphene on 5QL $Sb_2Te_3$ in Fig. 3(c) clearly indicate the spin-Hall phase of graphene. The denisty plot of a state near the Dirac point in Fig. 3(d) highlights the coupling between graphene $p_z$ orbitals and TI surface states. The giant increase of the intrinsic spin-orbit interaction is a result of proximity of Dirac points of graphene and TI. The effective spin-orbit interaction in graphene on TI substrate can be written as [26] $\lambda_I \propto \frac{|V_{\pi S}|^2}{(\varepsilon_\pi - \varepsilon_S)^2} \lambda_S$ where $V_{\pi S}$ is the hopping matrix element between π band and the surface states, $\lambda_S$ is the spin-orbit interaction in $Sb_2Te_3$, $\varepsilon_\pi$ and $\varepsilon_S$ are the energies of π and surface states, respectively. When the energy levels of the Dirac point in graphene and $Sb_2Te_3$ are very close to each other, we expect a resonance-type enhancement in the effective spin-orbit interaction in graphene. We found that such giant enhancement does not occur when the energies of Dirac points of graphene and TI are separated large.

The signature of the graphene SOC enhancement can be measured by various experimental techniques. The SOC splitting will produce the van Hove singularity (vHS) in the density of states (DOS), and the spin-polarized scanning tunneling spectroscopy (STS) can probe such sharp peaks in the DOS. Figure 4(a) shows our first-principle calculations of the DOS of graphene on 4QL $Sb_2Te_3$ slab with ($x$=1) or without SOC ($x$=0) in the Hamiltonian to resolve the features driven by the TI surface states. The vHS at about ~0.05 eV below and above the Fermi level is due to the spin-orbit gap in the graphene. Grown TI substrates have a large variation in potential profile [27], and thus graphene on TI substrates is expected to exhibit domains of the spin-Hall phase. Such puddles of spin-Hall phase in graphene as illustrated in Fig. 4(b) can also be measured by STS. Another straightforward way to detect the helical states is to study graphene nanoribbons or edges on $Sb_2Te_3$. Different from isolated graphene nanoribbons, which have a very small spin gap for particular edge atomic structures in case of very narrow width [28], graphene edges on top of TI will have spin-polarized conducting channels protected from atomic irregularities regardless of the width.

In summary, we studied the electronic structure of epitaxial graphene on top of $Sb_2Te_3$ topological insulator using first-principles calculations and tight-binding methods. We showed that a giant spin-orbit interaction of three orders of magnitude larger than the intrinsic value of graphene is induced in the epitaxial graphene so that it turns into the spin-Hall phase. This large enhancement of the spin-orbit interaction in graphene was found to be not simply because graphene is close to the surface of topological insulator but rather due to the proximity of graphene Dirac cones to that of topological insulator. Our results demonstrate that graphene can not only be used as a probe of TI surface states but also work as fascinating spin transport structures in combination with topological insulators.

ACKNOWLEGMENT

We would like thank Steven G. Louie and Marvin L. Cohen for helpful discussion. This works was supported by the National Research Foundation of Korea (NRF) grant funded by the Korea government (MEST) (Grant No. 2009-0087731, SRC program No. 2011-0030046 and WCU program No. R31-2008-000-10059). The authors would like to acknowledge the support from KISTI supercomputing center through the strategic support program for the supercomputing application research (No. KSC-2010-C2-0008). SHJ also acknowledges the support from National Science Foundation Grant No. DMR10-1006184 and by the Director, Office of Science, Office of Basic Energy Sciences, Materials Sciences and Engineering Division, U.S. Department of Energy under Contract No. DE- AC02-05CH11231.

FIGURES

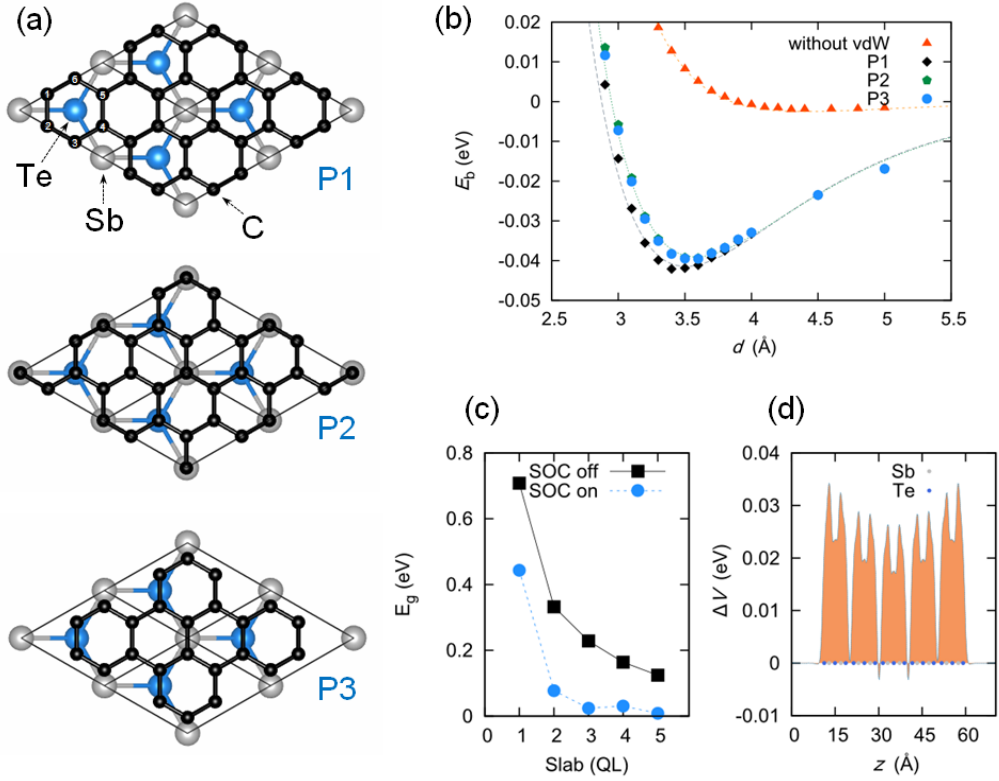

Figure 1 (Color online). (a) Top views of atomic structures of epitaxial graphene on (111) surface of $Sb_2Te_3$ thin film (slab) modeled by $\sqrt{3}\times\sqrt{3}$ R30° supercell. Three contact configurations from top to bottom: P1, carbon hexagon centers on top of surface Te atoms (blue balls); P2, carbon atoms (grey balls) on top of surface Te atoms; P3, carbon-carbon bridges on top of surface Te atoms. (b) Calculated binding energy curves of graphene on $Sb_2Te_3$ with vdW interactions included as a function of binding distance ($d$). Without van der Waals corrections, GGA cannot describe the binding correctly. (c) Calculated (indirect) band gaps of $Sb_2Te_3$ slabs using first-principles methods including SOC (blue circles) and without it (filled boxes) as a function of slab thickness. (d) The differences in electrostatic potentials ($\Delta V=V_{so}-V_{sp}$) with and without SOC from our first-principles calculations of 5QL $Sb_2Te_3$ slab, where $V_{so}$ and $V_{sp}$ are the Hartree potentials including SOC and including only spin-polarization without SOC, respectively.

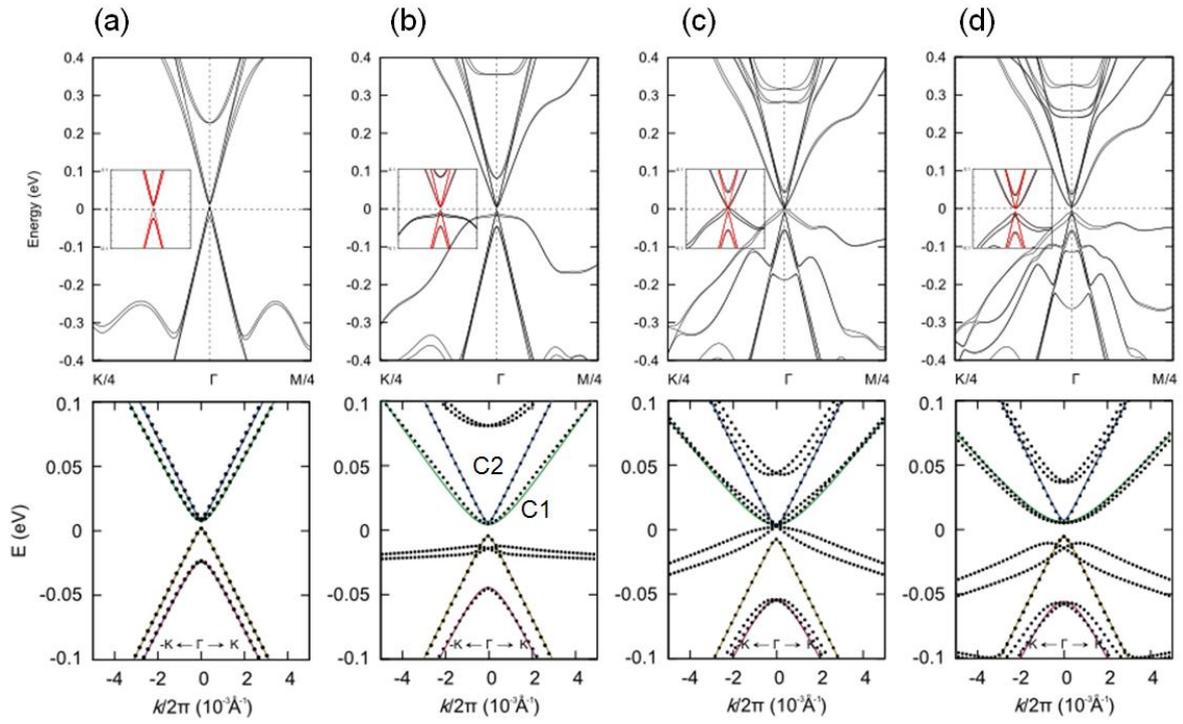

Figure 2 (Color online). Band structures of epitaxial graphene on top of $Sb_2Te_3$ slab (a) 1QL, (b) 2QL, (c) 3QL and (d) 4QL. The insets in the upper panels that detail the electronic states from graphene (red lines) and from TI (black lines) near the Fermi level (zero energy), are enlarged in the lower panels with dots representing first-principles calculations and lines the fitting to the tight-binding Hamiltonian for graphene.

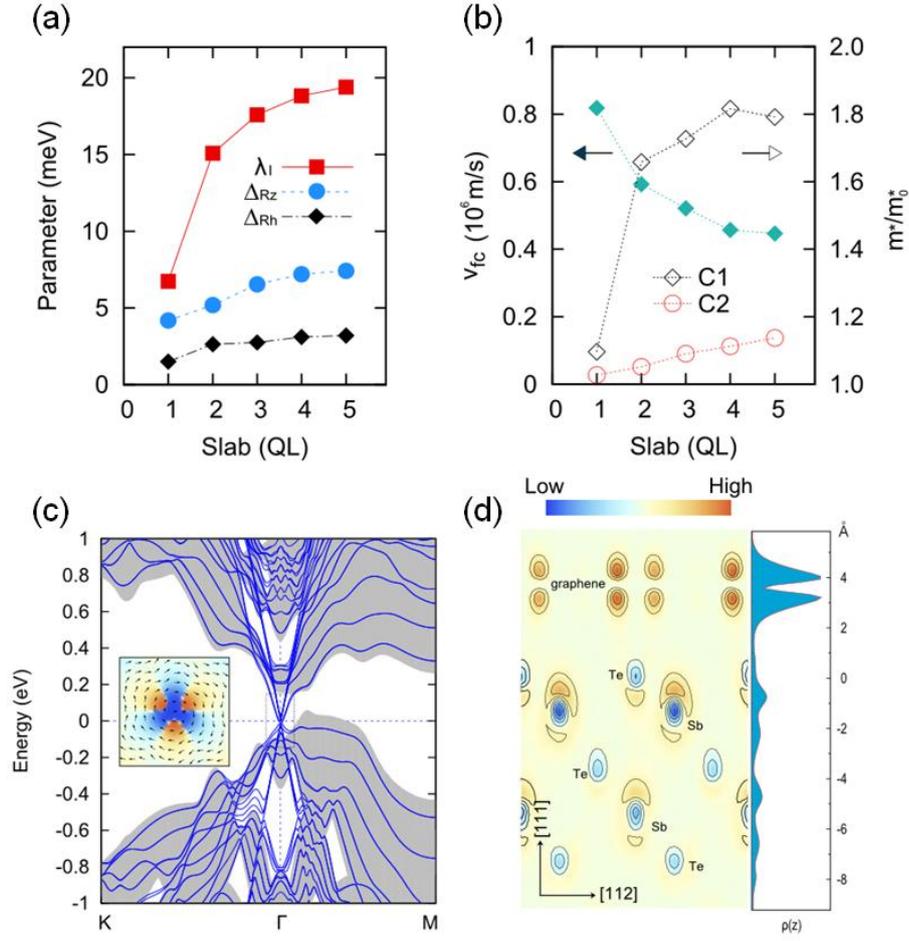

Figure 3 (Color online). (a) Calculated results of SOC strength of graphene on top of $Sb_2Te_3$ slabs of 1~5 QL after fitting the first-principles calculations to the tight-binding Hamiltonian. $\lambda_I$, intrinsic SOC; $\Delta_{Rz}$ and $\Delta_{Rh}$, Rashba SOC due to normal electric field and in-plane electric field, respectively. (b) Dependence on $Sb_2Te_3$ substrate thickness of Fermi velocity [filled diamonds for C1 band in Fig. 2(b)] and the cyclotron mass (open diamonds and circles for the conduction bands C1 and C2, respectively, normalized to that of pristine graphene) of graphene Dirac cones. (c) Calculated band structure of epitaxial graphene on 5QL $Sb_2Te_3$ slab (in blue lines) superimposed with the bulk band structures projected onto the surface (shaded areas). The inset is the spin helical structure of Dirac fermions in graphene. (d) The squared wave function of a state near the Dirac point in the (112) plane and its integrated charge density $\rho(z)$ along [111] direction (right panel). The surface atomic layer of $Sb_2Te_3$ is at z=0.

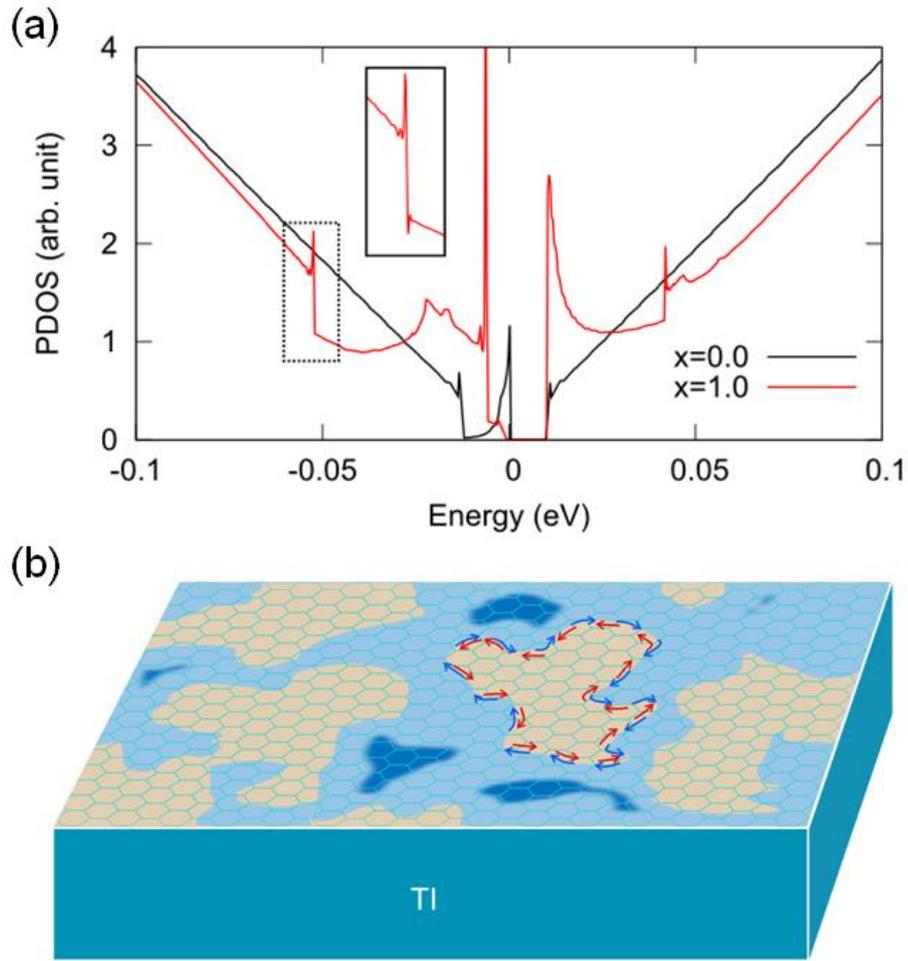

Figure 4 (Color online). (a) First-principle calculations of density of states (DOS) of graphene on 4 QL $Sb_2Te_3$ with ($x$=1) or without SOC ($x$=0). The energy gap at the Fermi level is due to the change in the hopping parameter between next-nearest carbon atoms. We observe the van Hove singularities at around ~0.05 eV below and above the Fermi level, which are due to the SOC in graphene, as well as at the energy-gap edges. (b) Schematic view of the spin-polarized edge states at the phase boundary between normal and spin-Hall phases in graphene. Due to the local variation in chemical potential in $Sb_2Te_3$ surface, we expect the puddles of spin-Hall phase in graphene.